\documentclass{article}
\usepackage{amsfonts}
\usepackage{amsmath}
\usepackage{amssymb}

\usepackage{alltt}
\usepackage{psfig}

\begin{document}

\title{Direct Test of the Time-Independence of Fundamental Nuclear Constants
Using the Oklo Natural Reactor \footnote{Lecture given at ATOMKI, 18 November, 1982.  This document was re-entered 
manually; scanned original is available at http://alexonline.info .  
This document was uploaded to arXiv.org by Ilya Shlyakhter (contact info at http://ilya.cc)
after the death of its author.}}

\author{Alexander I. Shlyakhter}

\date{November 18, 1982}

\maketitle

% command to enter a chemical isotope: \isot{O}{8}{16}
\newcommand{\isot}[3]{\ensuremath{^{#2}_{#3}\textrm{#1}}}
\newcommand{\iso}[2]{\ensuremath{^{#2}\textrm{#1}}}
\newcommand{\e}[2]{\ensuremath{{#1} \cdot 10^{#2}}}
\newcommand{\ee}[1]{\ensuremath{10^{#1}}}

\section{Introduction}

The following eight quantities enter the basic laws of physics and are generally
regarded as the ``fundamental constants'' (I follow Dyson's review \cite{1} in this section).

\begin{enumerate}
\item $c = 3 \cdot 10^{10} cm \cdot sec^{-1}$, velocity of light
\item $\hbar = 1.05 \cdot 10^{-27} erg \cdot sec$, Planck's constant
\item $e = 4.8 \cdot 10^{-10} erg^{1/2}cm^{1/2}$, elementary charge
\item $m_p = 1.6 \cdot 10^{-24} gram$, mass of the proton
\item $g = 1.4 \cdot 10^{-49} erg \cdot cm^3$, Fermi's constant of weak interactions
\item $G = 6.7 \cdot 10^{-8} erg \cdot cm \cdot gram^{-2}$, constant of gravitation
\item $H = 1.6 \cdot 10^{-18} sec^{-1}$, Hubble's constant ($1/H \approx 2 \cdot 10^{10}$
years gives the ``age'' of the Universe)
\item $\rho = 10^{-31} gram \cdot cm^{-3}$, mean density of mass in the Universe
\end{enumerate}

This list is not exhaustive, of course.  In particular, the constants of strong
interactions could be added to it.

The ``cosmological'' quantities $H$ and $\rho$, which refer to the Universe as a whole,
vary as it expands: they are decreasing at the rate of about $5 \cdot 10^{-11} yr^{-1}$.
On the other hand, the ``laboratory'' quantities 1)-6) are generally believed to be
exactly constant.  Milne \cite{2} and Dirac \cite{3} were the first to notice that this is no 
more than a hypothesis, requiring experimental confirmation.

I'd like to emphasize that only those variations of fundamental constants which change at
least one dimensionless ratio of the dimensional quantities have physical meaning.
The reason is that only such ratios do not depend on the choice of units and standards.

From the dimensional quantities 1) - 8) five dimensionless ratios can be formed:

\begin{enumerate}
\item $\alpha = e^2 / \hbar c \approx 1/137$
\item $\beta = (g m_p^2 c)/\hbar^3 = 9 \cdot 10^{-6}$
\item $\gamma = (G m_p^2)/(\hbar c) = 5 \cdot 10^{-39}$
\item $\delta = (H \hbar) / (m_p c^2) = 10^{-42}$
\item $\epsilon = (G \rho) / H^2 = 2 \cdot 10^{-3}$
\end{enumerate}

Note that the inverse of $\delta$ gives $\delta^{-1} = 10^{42}$ which is the age
of the Universe measured in ``tempons'' (atomic units of time).

According to the conventional view the ratios of the ``laboratory'' quantities
$\alpha$, $\beta$, and $\gamma$ did not change their numerical values during
the 20 billion years since the ``Big Bang''.

The Conventional View: $\alpha, \beta, \gamma = const; \beta \sim t^{-1}, \gamma \sim t^{-1}$.

Dirac \cite{3} introduced the ``Numerological Principle'' (or ``The Large Numbers Hypothesis''
(LNH) which states that ``all very large dimensionless numbers which can be constructed
from the important natural constants of cosmology and atomic theory are connected by simple
mathematical relations involving coefficients of the order of magnitude unity'' [3b].
For example, the large value of $\gamma^{-1}$ could prove compatible with the numerological
principle if it was proportional to $\delta^{-1}$ and thus was time-dependent.

The majority of the proposed versions of the possible variation of constants is based on
similar arguments.  Three of them are listed below (see \cite{1,3,4,5}).
\begin{itemize}
\item Dirac (1937): $\alpha, \beta, \epsilon = const; \gamma \sim t^{-1}, \delta \sim t^{-1}$
\item Teller (1948): $\beta, \epsilon = const, \alpha^{-1} \sim \ln(\gamma^-1), \gamma \sim t^{-1}, \delta \sim t^{-1}$
\item Gamow (1967): $\beta,\gamma,\epsilon = const, \alpha \sim t, \delta \sim t^{-1}$
\end{itemize}

These versions predict the rate of variation of constants at the present epoch about
$10^{-10} - 10^{-12} yr^{-1}$.

\section{Experimental Limits on the Rate of Variation of ``Nuclear'' Constants}

(The limits of the possible change of the constant of gravitation are discussed in 
\cite{1,6,7,25}).

Direct experimental evidence comes either from astrophysics or from geophysics.  Astrophysical
data allow judgement on the large-scale uniformity of physical laws in space (at distances
up to 15 billion light years).  Geophysical ones provide evidence on the absence of
variation of constants along the world-line of the Earth since its crust became solid
($\approx$ 4.5 billion years ago).

The data on the absorption spectra of the distant quasars show that the numerical value
of the dimensionless quantity $|\alpha^2 \cdot g_p \cdot m_e/m_p|$ is the same throughout
the observable Universe with the accuracy of about $10^{-4}$ \cite{8}.  If one assumes the
Friedman model, this limit restricts the possible rate of variation of $\alpha$ by
$\approx 10^{-14} yr^{-1}$.

The decay rate $\lambda$ of radioactive nuclide depends on nuclear constants.  For example,
in the case of high $Z$ and small decay energy $\delta$ the $\beta$-decay rate $\lambda_\beta$
is highly sensitive to the value of $\alpha$.  The estimate of the ``sensitivity'' $s$
gives \cite{1}

\[ s \equiv \frac{\delta \lambda_\beta}{\lambda_\beta} \Big{/} \frac{\delta \alpha}{\alpha} = -(2Z+1)(2+\sqrt{1-\alpha^2 Z^2})
\cdot A^{-1/3} \cdot 0.6 [MeV/\Delta] \]

For the transition $\isot{Re}{187}{75} \rightarrow \isot{Os}{187}{76}$ 
($\textrm{T}_{1/2} \approx 40$ billion years, $\Delta = 2.5$ keV), this estimate gives
$s = -2 \cdot 10^4$.  Using the data on the abundances of rhenium and osmium
isotopes, Dyson \cite{1} obtained the following upper limit on the rates of variation
of $\alpha$ and $\beta$:
\[ \textrm{Dyson (1972): } \Big{|} \frac{\dot{\alpha}}{\alpha} \Big{|}
 \le 2 \cdot 10^{-14} yr^{-1},
\Big{|} \frac{\dot{\beta}}{\beta} \Big{|} \le 10^{-10} yr^{-1} \]

If one assumes that $\beta$ does not change with time, the limit for $\alpha$ is:
\[|\dot{\alpha} / \alpha| \le 5 \cdot 10^{-15} yr^{-1}\].

\section{Neutron Resonances as the Sensitive Indicators of the Variation of Nuclear Constants}

Several years ago I noticed that because of the sharp resonances in its absorption
cross-section, the heavy nucleus is a highly tuned detector of neutrons.  Resonances
will shift along the energy scale if there is a change in the nuclear potential,
by analogy with the shift in the reception frequency in an ordinary radio receiver
when there is a change in the parameters of the resonance circuit \cite{9}.

\begin{figure*}[tbph]
\centerline{\psfig{figure=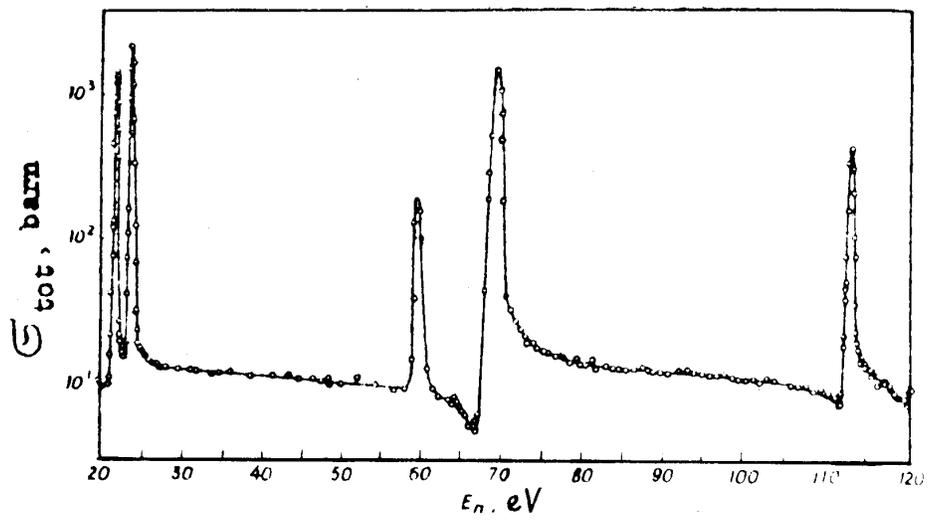,width=5in}}
\caption{Resonances in the energy dependence of the total neutron cross section of \iso{Th}{232} \cite{10}}
\label{fig:fig1}
\end{figure*}

\begin{figure*}[tbph]
\centerline{\psfig{figure=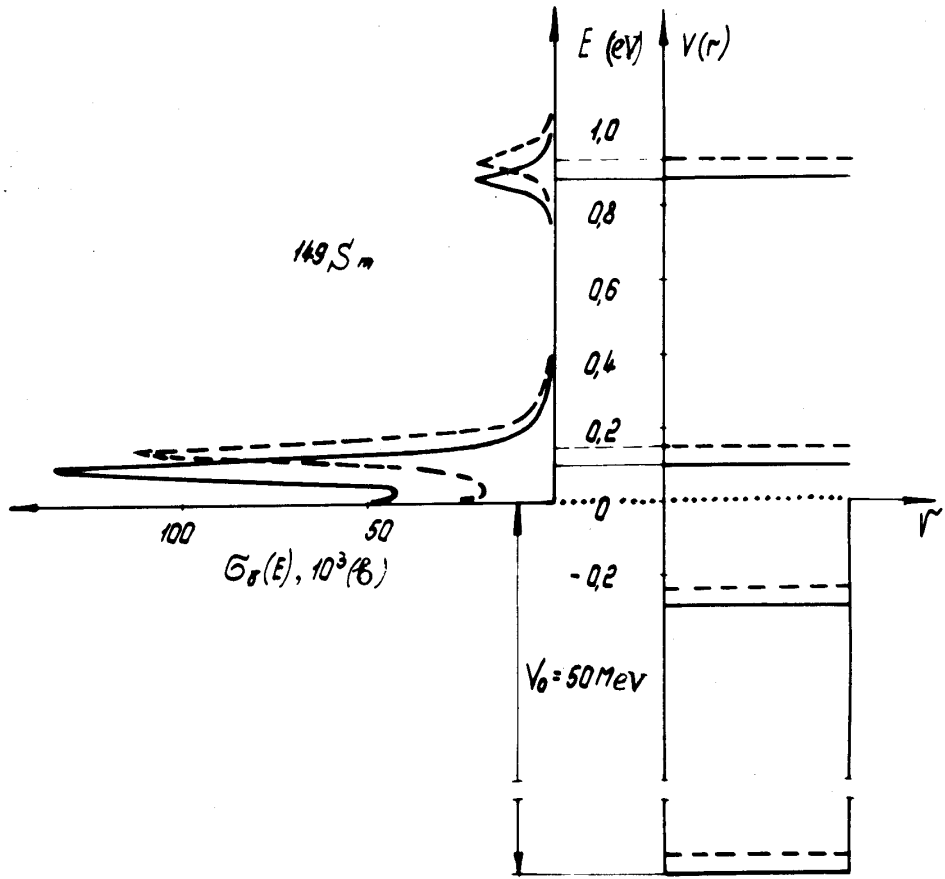,width=5in}}
\caption{Two energy scales in the nucleus: the eV scale of neutron resonances and MeV scale of the potential well.  Solid lines show actual positions of resonances and the energy dependence of capture cross section for \iso{Sm}{149}.
 The dashed ones demonstrate the effect of the variation of nuclear constants. }
\label{fig:fig2}
\end{figure*}

For the incident neutron, the nucleus presents a potential well with the depth of
about $V_0 = 50$ MeV.  At low neutron energy the cross-section exhibits sharp
resonances (Fig. \ref{fig:fig1}).  Their positions are measured with the accuracy 
$\Delta_{exp} \sim 10^{-2}$ eV.  Thus, there are two energy scales: $V_0$
and $\Delta_{exp}$.  Any change of $V_0$ by $\Delta V_0$ would cause the shift
of all nuclear levels including the levels of compound nucleus, i.e. neutron
resonances (Fig. \ref{fig:fig2}).

The dimensionless quantity entering this problem is the ratio of the depth of the
potential well $V_0$ to the uncertainty in the resonance energy $\Delta_{exp}$.
This suggests that variations of the basic nuclear parameters are amplified
in the shift of resonances by an enormous factor $s \sim 10^{10}$.

Unfortunately, it seems very difficult to calculate consistently the shift
of a given neutron resonance caused by the variation of the fundamental
nuclear constants.  Here I shall use the simplest assumption that neutron
resonances are shifted by $\Delta V_0$ like single-particle levels in a 
potential well.  Then the experimental evidence showing that the shift
of the resonances during the time period $T$ has not exceeded $\Delta_{exp}$
imposes the following limits on the possible variation of the interaction constants:
\begin{itemize}
\item strong: $|\dot{V_0}/V_0| \le \Delta_{exp} / (V_0 \cdot T) = 2 \cdot 10^{-8} \cdot
 \Delta_{exp} (eV)/T(yrs)$
\item electromagnetic: $|\dot{\alpha}/\alpha| \sim 20|\dot{V_0}/V_0|$
\item weak: $|\dot{\beta}/\beta| \sim 5 \cdot 10^6 |\dot{V_0}/V_0|$
\end{itemize}
I follow Gamow [5b] in assuming that the variation of the strong interaction constants
is adequately reproduced by the change in the depth of the nuclear potential well.
The estimate for $|\dot{\alpha}/\alpha|$ is based on the equation of nuclear
compressibility \cite{10}.  For nuclei with $A \sim 150$ the change in the radius appears
to be 40 times less than the change in $\alpha$ and the corresponding shift of the
levels is 20 times less.  The limit for $|\dot{\beta}/\beta|$ is obtained assuming
that the contribution of weak interactions to the nuclear binding energy
is of the order of $2 \cdot 10^{-7}$ \cite{11,12}.

These estimates demonstrate that if there existed a Precambrian physicist
who could measure the energies of the neutron resonances about a billion years ago
with an accuracy about $10^{-2}$ eV then the limits of the possible variation of
the fundamental nuclear constants could be improved by several orders of magnitude.
The striking discovery of the ``Oklo natural nuclear reactor'' proves that such
seemingly improbable experiment has been actually performed nearly 2 billion
years ago and the results have been reliably ``recorded'' in the isotopic
composition of the elements in the reactor core.

\section{The Oklo Phenomenon}

Oklo is the name of a locality in the Gabon Republic (West Africa) where
the open-pit uranium mine is situated.  About 1.8 billion years ago
within a rich vein of uranium ore ``the natural reactor'' went critical,
consumed a portion of its fuel and then shut down.  The total amount
of energy produced by the reactor is estimated as $1.5 \cdot 10^4$
megawatt years which seems to be enough for a city like Budapest
for about a year.  You now may have a question: how could it appear that
even in the design of nuclear reactors, which is generally considered
to be one of the most impressive achievements of science and technology of
our century, ``the man was not an innovator but an unwitting imitator
of nature'' \cite{17}.

In fact, no natural reactor could operate today with uranium containing
only 0.72 percent of the fissile isotope \iso{U}{235}.  The ratio of \iso{U}{235}
to \iso{U}{238}, however, has not been constant throughout the history of the
Earth.  The half-life of \iso{U}{235} is about 700 million years, that of
\iso{U}{238} about 4.6 billion years.  Thus, 2 billion years ago the abundance
of \iso{U}{235} was about 3 percent (note that in the contemporary power-producing
reactors uranium is enriched up to the same value).  In 1956 Kuroda \cite{13} showed
that at that epoch under favourable conditions (i.e. concentration of water
must be high enough and that of the strong absorbers of neutrons low enough)
the spontaneous nuclear chain reaction could take place in rich uranium deposits.
However, until 1972 no traces of a natural reactor have been found.

In June 1972 the uranium slightly depleted in \iso{U}{235} was first detected
at a French uranium-enrichment plant.  The anomaly was traced through the
numerous stages of the manufacturing process right back to the ore-enriching
plant at Mounana near Franceville in Gabon.  The original ore with mean \iso{U}{235}
abundance of 0.4-0.5\% was mined at Oklo.  The French Atomic Energy Comission (CEA)
then initiated the investigation of this ``Oklo Phenomenon''.  The analysis of
the numerous samples obtained by drilling revealed the distribution of isotopic
anomalies in detail.  The results were discussed at the IAEA symposiums \cite{14,15}
and reviewed in \cite{16,17,18,19}.

\begin{figure*}[tbph]
\centerline{\psfig{figure=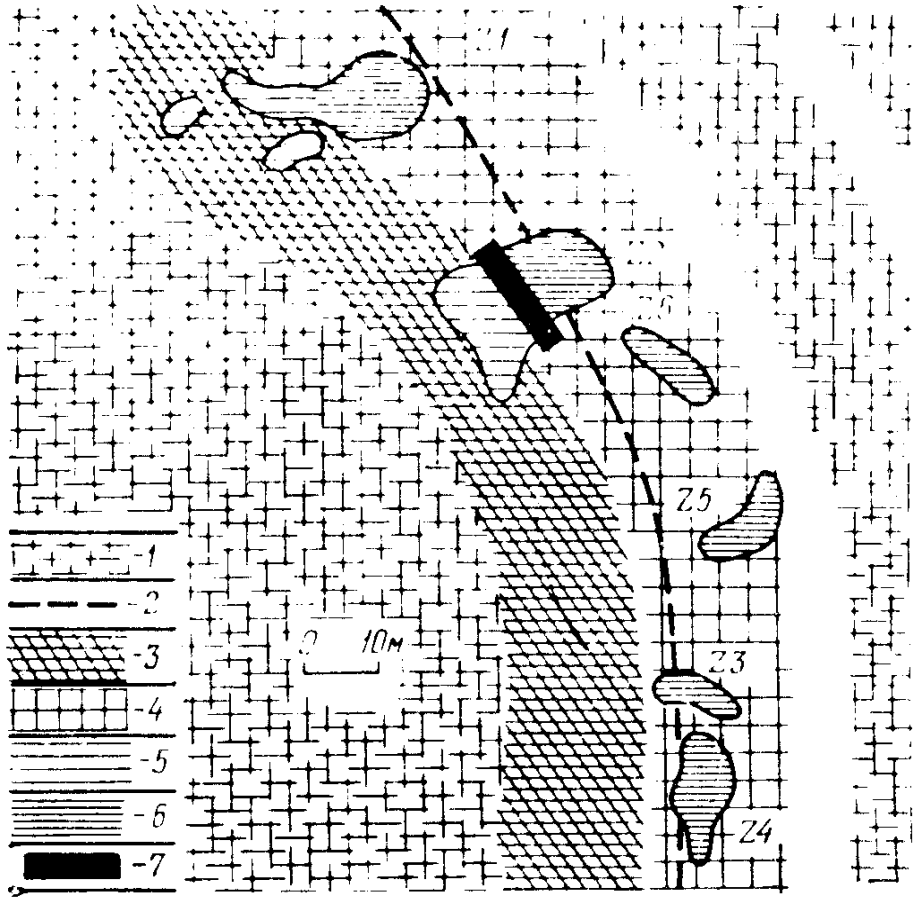,width=5in}}
\caption{Disposition of the active zones in the Oklo reactor [18]: 1 - sandstone; 2 - boundary of mined ore;
 3 - sandstone wall; 4 - floor of pit; 5 - mined part of reactor; 6 - explored part of reactor; 7 - area to be
 preserved for future studies.}
\label{fig:fig3}
\end{figure*}

\begin{figure*}[tbph]
\centerline{\psfig{figure=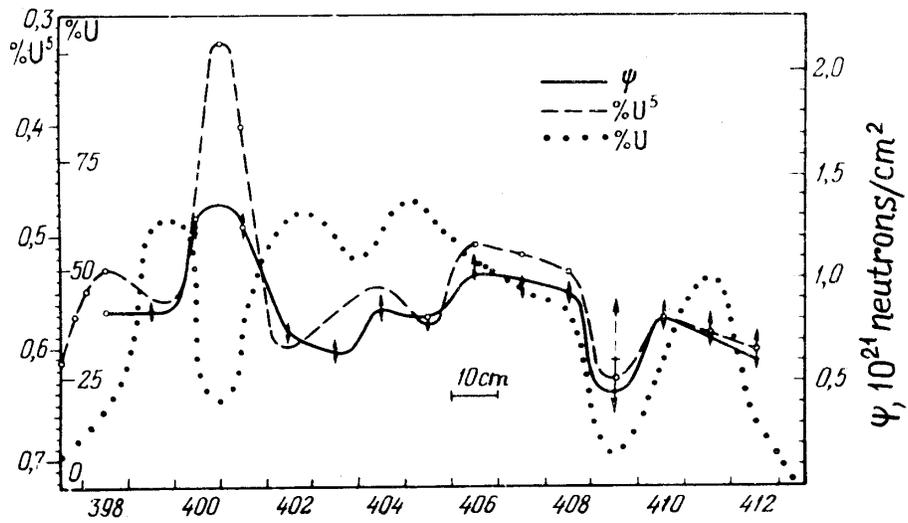,width=5in}}
\caption{Distribution of the integrated neutron flux $\psi$, \iso{U}{235} concentration in uranium and the
 concentration of uranium in the ore.  Sample numbers are plotted along the abscissa axis. \cite{18} }
\label{fig:fig4}
\end{figure*}

\begin{figure*}[tbph]
\centerline{\psfig{figure=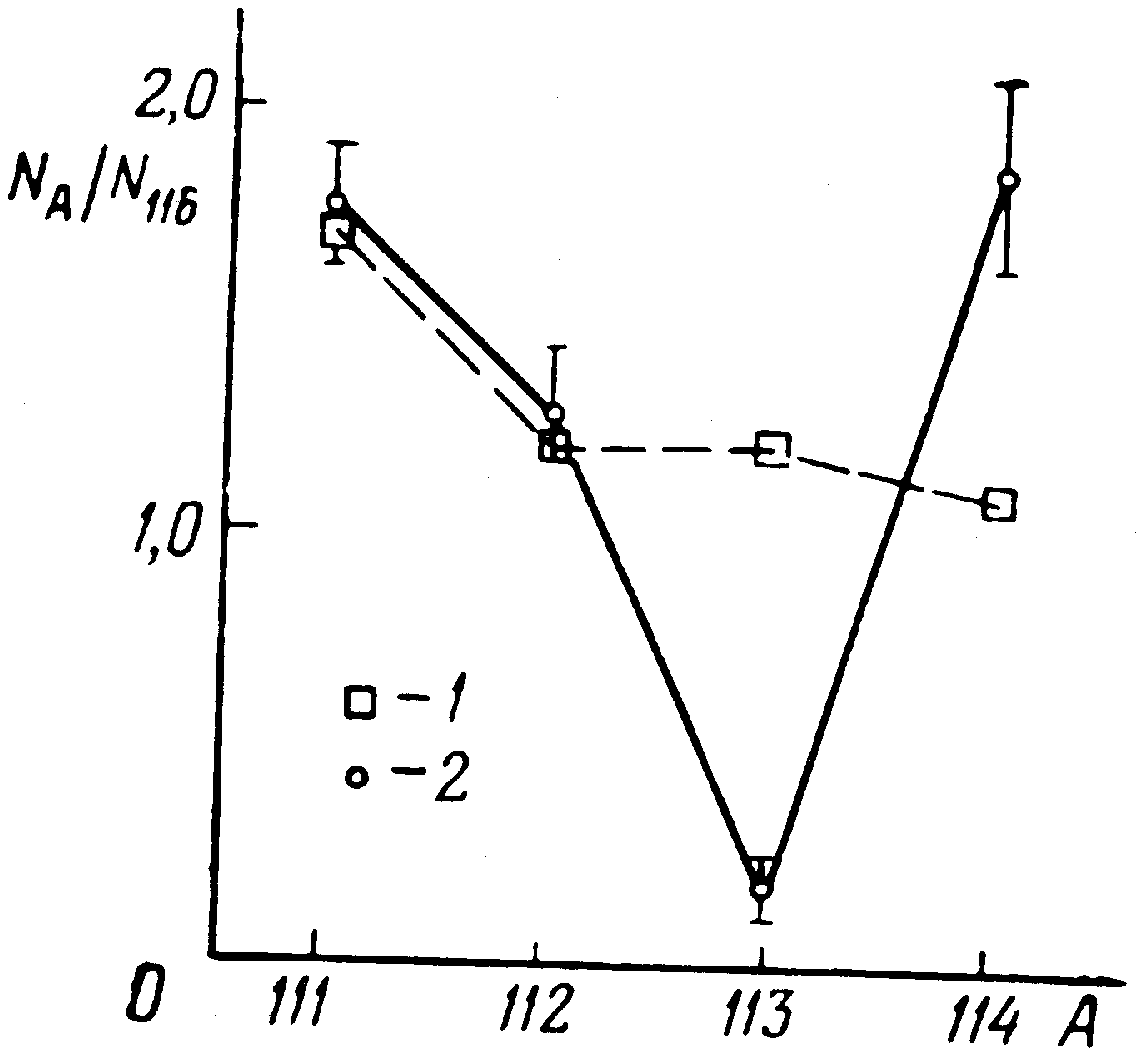,width=5in}}
\caption{Burn-up of \iso{Cd}{113} in the Oklo reactor: 1 - yield of Cd isotopes relative to \iso{Cd}{116}
 during fission; 2 - data for the Oklo sample \cite{18,20}.}
\label{fig:fig5}
\end{figure*}

Altogether 6 reaction zones (Fig. \ref{fig:fig3}) were found over a distance of a few dozen meters.
The spectrum of fission products (especially of the rare-earths) gives the quantitative
evidence that a natural reactor once operated there.  Both the absolute amounts
of these elements and their isotopic composition can be explained only by their
origin in fission (Table \ref{tab:tab1}).  The fluence (the flux integrated over time) of thermal
neutrons, which can be determined e.g. from the \iso{Nd}{144}/\iso{Nd}{143} ratio,
reaches the value of $1.5 \cdot 10^{21} n/cm^2$ (Fig. \ref{fig:fig4}).  This means that
the strong absorbers of thermal neutrons (having capture cross sections 
$\sigma_{\gamma} \gtrsim 2 \cdot 10^3$ barn) must become heavily depleted.  At the same
time the concentration of the next (in A) isotope will increase.  This phenomenon
has been observed experimentally (Fig. \ref{fig:fig5}).  This in itself shows immediately
that a chain reaction initiated by thermal neutrons has taken place.

\section{The ``Measurement'' of the Energies of Neutron Resonances 2 Billion Years Ago
Using the Oklo Data}

In the strong absorbers of thermal neutrons their large capture cross sections are
determined in each case by a single resonance which is occasionally located near
zero neutron energy.  The cross section is given by the Breit-Wigner formula
\[ \sigma_{\gamma} = g \cdot \pi \lambda^2 \cdot \frac{\Gamma_n \cdot \Gamma_{\gamma}} 
{(E-E_r)^2 + (\Gamma/2)^2} \]

\begin{figure*}[tbph]
\centerline{\psfig{figure=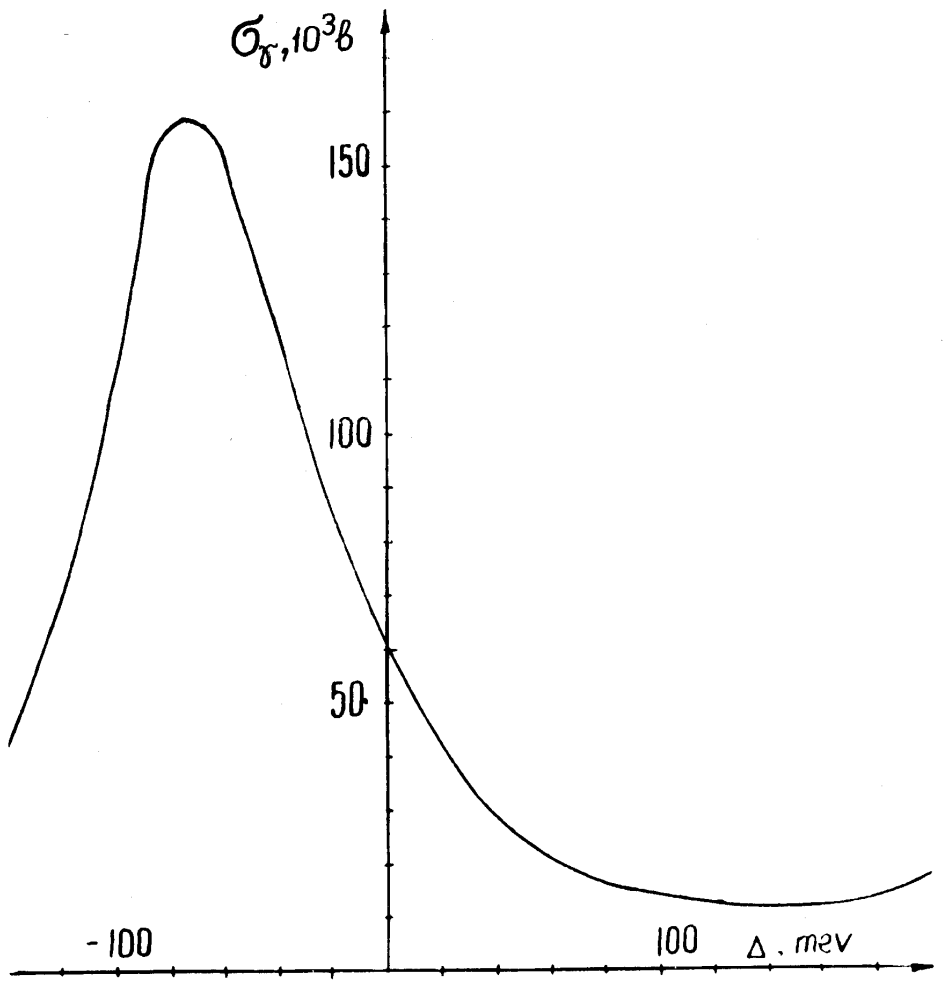,width=5in}}
\caption{The variation of the \iso{Sm}{149} capture cross section (averaged over the maxwellian spectrum
 of neutrons with kT = 0.025 eV) when the resonances are shifted by $\Delta$ ([9a]).}
\label{fig:fig6}
\end{figure*}

Here $g$ is the statistical factor, $E$ is the neutron energy and $\lambda$
is the corresponding wavelength, $\Gamma_n$ and $\Gamma_\gamma$ are the
partial elastic and capture widths, and $\Gamma$ is the total width of the
resonance.  The cross section changes sharply when the resonance is shifted
along the energy scale.  Fig. \ref{fig:fig6} shows this effect for the maxwellian-averaged
(at kT=0.025 eV) capture cross section of \iso{Sm}{149}.  From the relative
concentrations of samarium isotopes and the neutron fluence independently
determined at the same points of the reactor one can extract the value of the
capture cross section at the epoch of chain reaction.  For example, R. Naudet et
al \cite{14} have measured the isotopic composition of uranium, neodymium and samarium
in 50 samples and have managed to determine the fluence $\psi$ from U and Nd data
reliably for 36 of them.  Then the cross section of \iso{Sm}{149} is given
by the following equation
\[ \frac{N_{147}+N_{148}}{N_{149}} = \frac{\gamma_{147}}{\gamma_{149}} \cdot
\frac{\bar{N}_{235}}{N_{235}} \cdot \sigma_{149} \]

Here $N_i$ denotes the final concentrations of samarium isotopes, $\bar{N}_{235}/N_{235}$
is the ratio of the average \iso{U}{235} concentration during the period of the reaction
to its final value, $\gamma_i$ are the yields of \iso{Sm}{147,149} from the fission of
\iso{U}{235} ($\gamma_{148}$ is negligible).

The analysis of samarium data for the same 36 samples where $\psi$ is known gives
the value of $\sigma_{149}^{Oklo}$ ``measured'' 2 billion years ago

\[ \sigma_{149}^{Oklo} = (55 \pm 8) \cdot 10^3 \textrm{\ barn} \]

the contemporary value being $\approx 60 \cdot 10^3$ barn (this value depends on
the spectrum of thermal neutrons, here I use the maxwellian spectrum at
$T = 300^\circ K$).  Taking into account two standard erros, we obtain [9a]

\[ |\Delta_{exp}| \le 20 \cdot 10^{-3} eV \]

If one takes into account also the data for europium (which are less precise)
with three standard errors the result is [9b]

\[ |\Delta_{exp}| \le 50 \cdot 10^{-3} eV \]

Note that the effect of such a small shift of resonances upon the capture cross sections
of uranium and neodimium is negligible, so the fluence $\psi$ is determined reliably.

Yu.V.Petrov \cite{18} has pointed out that one could avoid determining $\psi$ if the
relative concentrations of two strong absorbers were available.  In this case the
absence of a shift in the resonance of one absorber relative to the other can
be verified directly.

\begin{figure*}[tbph]
\centerline{\psfig{figure=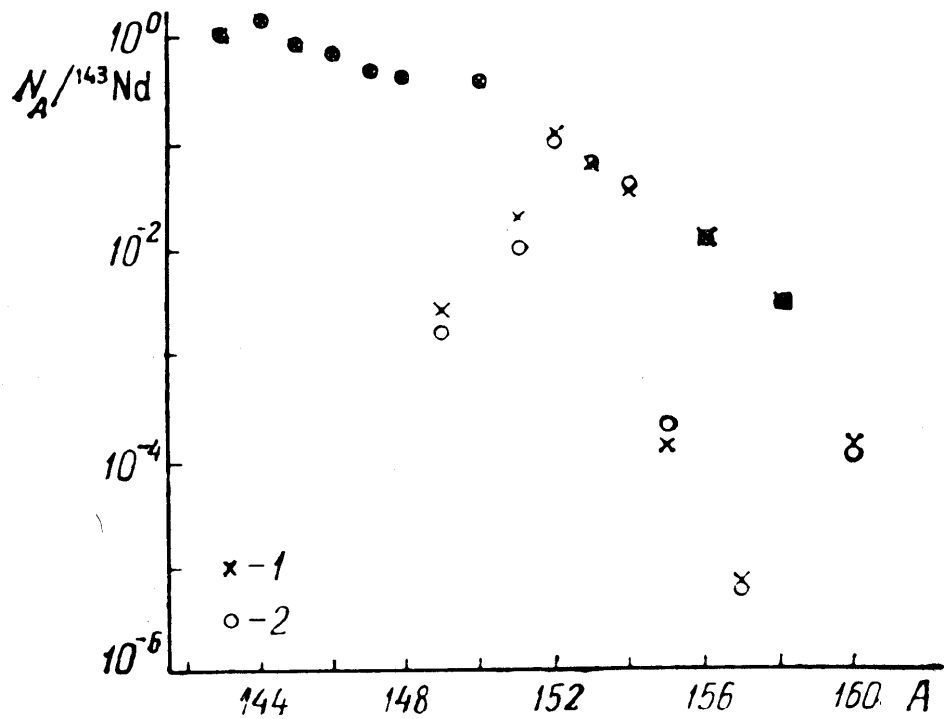,width=5in}}
\caption{Comparison of calculated (1) and measured (2) concentrations of fission products relative to the
 \iso{Nd}{143} concentration for one of the Oklo samples. \cite{20}}
\label{fig:fig7}
\end{figure*}

The absence of an appreciable shift of near-threshold resonances also follows qualitatively
from the fact that all the contemporary strong absorbers were strongly burnt up
in the Oklo reactor, whereas the weak absorbers were weakly burnt up \cite{18}.  In addition
to the cadmium data (Fig. \ref{fig:fig5}) the results of the measurements of the conentration of rare-earth
elements relative to \iso{Nd}{143} in one of the Oklo samples \cite{20} are reproduced in Fig. \ref{fig:fig7}.
The dips in the distribution correspond to strong absorbers: \iso{Sm}{149}, \iso{Eu}{151},
\iso{Gd}{155} and \iso{Gd}{157}.  The burn-up depth, calculated using the
contemporary values of absorption cross section is in excellent agreement with experiment,
especially if we recall that the neutron spectrum over which the cross section has
to be averaged is now known well enough.

We therefore conclude once again that, over the 1.8 billion years since the operation of the
Oklo reactor, the resonances or, in other words, the compound-nucleus levels,
have shifted by less than $\Gamma/2 \sim 50 \cdot 10^{-3}$ eV, i.e. the mean rate
of the shift did not exceed $3 \cdot 10^{-11}$ eV / year.  This is by three orders
of magnitude less than the experimental limit on the rate of change in the transition energy
in the decay of \iso{Re}{187}.  Unfortunately, at present there are no consistent
calculations that would have connected the position of each neutron resonance with the
nuclear potential parameters reliably.  However, even the preliminary estimates of Sec. 3
can be used to improve the limits obtained by other authors substantially (Table \ref{tab:tab2}).
These estimates evidently rule out a power law or a logarithmic asymptotic dependence
of the strong and electromagnetic interaction constants on the lifetime of the Universe.

\section{The Probability of an Occasional Coincidence}

I have assumed above that the variation of nuclear constants (if any) has been very small
so that the shift of resonances would have appeared much less than their average
separation.  One could imagine, however, a case in which even after a considerable variation
of the constants all the strong absorbers would have remained strong.  This could occur
if some other resonance appeared near the threshold and dominated in the capture
cross section.  In this section I shall estimate the probability of such a coincidence
using the recently developed statistical approach to estimating unknown thermal cross
sections \cite{21}.

For each nuclide one calculates the ``expected'' capture cross section $\sigma_\gamma^*$
using the average values of its resonance parameters.  The universal distribution function
$S_\gamma(z)$ has been calculated using the generally accepted distribution laws for
these parameters.  It gives the probability for the ratio of the actual cross section
$\sigma_\gamma$ to its expected value $\sigma_\gamma / \sigma_\gamma^*$ not to exceed $z$.

Table \ref{tab:tab3} which is taken from \cite{23} gives the probability for each strong absorber to remain
strong after a large variation of constants.  I assumed that its new cross section will
be at least half of its old value.  Those nuclides for which this probability is small appear
to be sensitive ``indicators'' of the variation of constants (e.g. \iso{Cd}{113} and
\iso{Gd}{157}).  On the other hand, \iso{Eu}{151} will remain a strong absorber with the
probability of about 0.3, thus being rather useless in this respect.  The product of
the values $1 - S_\gamma(z)$ for all nuclides gives the probability of a simultaneous
coincidence the estimate $P \sim 4 \cdot 10^{-7}$.  Note that this estimate is rather
conservative since if the resonances have shifted considerably some weak absorbers could
have been strong ones 2 billion years ago, giving rise to some mysterious isotopic anomalies
at Oklo none of which have been seen.

\section{Conclusions}

The analysis of the Oklo data provides very strong evidence in favour of the invariability 
of nuclear constants.  The shift of neutron resonances during the last 2 billion years
does not exceed $50 \cdot 10^{-3} eV$ or $3 \cdot 10^{-11} eV/yr$.  This is so far the
most precise limit and simple estimates of the rate of variation of the interaction
constants shown in Table \ref{tab:tab2} were cited on several occasions (see e.g. \cite{24,25,26,27}).
I must note, however, that these estimates should not be taken too seriously.
More accurate theoretical calculations of the influence of the fundamental
constants on the parameters of the neutron resonances are required.

On the other hand, Oklo is the only place on Earth where the variability of nuclear
constants (if any) could be detected.  For this reason it would be very interesting
to cary out special measurements in order to improve the limit $\Delta_{exp}$.

However, at present all the available data support the conventional view, according
to which the values of constants have not changed since the ``Big Bang''.  How could then
the ``Large Numbers'' coincidences be explained?  Zel'dovich \cite{28} has noted that within
modern quantum field theory, spontaneous topology change can readily give rise to
large numbers which are comparable to those considered by Dirac.  An alternative answer is
suggested by the so-called ``anthropic principle'' \cite{29,30,31} which states that only
those universes can ever become observable where the ``observers'' can survive.
The very possibility of life appears remarkably sensitive to the numerical values
of physical constants.  Following this line of argument all ``Large Numbers'' can be derived
without any appeal to the variation of constants.

It was a great honor for me to win the international competition of young scientists
in 1979 and to be invited to ATOMKI.  I would like to express here my deep gratitude
to V.A.Nazarenko and Yu.V.Petrov for their interest and support.

\clearpage

\begin{table}[tbph]
\label{tab:tab1}
\begin{center}
\begin{tabular}{|c|c|c|c|c|c|}
\hline
Isotopes of Nd              &   142   & 143+144 & 145+146 & 148  & 150  \\
\hline
Natural conentration, \%    &  27.11  &  36.02  &  25.52  & 5.73 & 5.62 \\
Fission of \iso{U}{235}, \% &    0    &  55.18  &  33.53  & 8.16 & 3.13 \\
\hline
Oklo samples, \%            &    0    &  54.95  &  33.46  & 8.25 & 3.34 \\
\hline
\end{tabular}
\caption{\cite{18} The agreement of the isotopic distribution of Nd with the fission
yields.  Fission products do not contain \iso{Nd}{142}, so that its amount was used to
determine the concentration of natural neodimium in the ore, and to introduce corrections
for it.}
\end{center}
\end{table}

\begin{table}[tbph]
\label{tab:tab2}
\begin{center}
\begin{tabular}{|c|c|c|}
\hline
Interaction                 &  Dyson, Davies \cite{1,22}   &  Present work  \\
\hline
strong, $yr^{-1}$           &  \e{2}{-12}             &   \e{5}{-19}   \\
\hline
electromagnetic,  $yr^{-1}$ &  \e{2}{-14}             &   $10^{-17}$   \\
\hline
weak, $yr^{-1}$             &  $10^{-10}$             &   \e{2}{-12}   \\
\hline
\end{tabular}
\caption{Comparison of upper bounds of the variation of nuclear constants. 
Estimates of Sec. 3 at $\Delta_{exp}$ = 50 meV and T = \e{1.8}{9}yrs are used.}
\end{center}
\end{table}

\begin{table}[tbph]
\label{tab:tab3}
\begin{center}
\begin{tabular}{|c|c|c|c|c|c|c|}
\hline
Nuclide  &  $\sigma_\gamma^*$, b & $\sigma_{\gamma\ exp}$, b & $z_{exp}$ & 
$\sigma_{\gamma\ min}^{Oklo}$ & $z_{min}^{Oklo}$ & $1 - S_\gamma (z_{min}^{Oklo})$ \\
\hline
\iso{Cd}{113} & 5.5  &  \e{(19.9 \pm 0.3)}{3} & \e{(3.6 \pm 0.9)}{3} & \ee{4} & \e{2}{3} & 0.01 \\
\iso{Sm}{149} & \e{5.6}{2} &  \e{(41 \pm 2)}{3} & 73 $\pm$ 16 & \e{2}{4} & 35 & 0.09 \\
\iso{Eu}{151} & \e{1.9}{3} &  \e{(9.2 \pm 2)}{3} & 4.8 $\pm$ 0.9 & \e{5}{3} & 2.6 & 0.33 \\
\iso{Gd}{155} & \e{5.7}{2} &  \e{(61.0 \pm 0.5)}{3} & 108 $\pm$ 18 & \e{3}{4} & 50 & 0.07 \\
\iso{Gd}{157} & \e{1.6}{2} &  \e{(254 \pm 2)}{3} & \e{(1.6 \pm 0.3)}{3} & \ee{5} & \e{6}{2} & 0.02 \\
\hline
\end{tabular}
\caption{The probability for a strong absorber to remain strong after a large
variation of constants \cite{23}.}
\end{center}
\end{table}

\end{document}